\documentclass[a4paper,11pt]{article}
\usepackage{pos}
\usepackage{microtype}

\let\oldbibliography\thebibliography
\renewcommand{\thebibliography}[1]{%
  \oldbibliography{#1}%
  \setlength{\itemsep}{0.05cm}%
}

\renewcommand{\logo}{\relax}

\title{Dijet photoproduction and transverse-plane geometry in ultra-peripheral nucleus-nucleus collisions}
\ShortTitle{Dijet photoproduction and transverse-plane geometry in UPCs}

\author{Kari\,J.\,Eskola}
\author{Vadim\,Guzey}
\author{Ilkka\,Helenius}
\author*{Petja\,Paakkinen}
\author{Hannu\,Paukkunen}

\affiliation{University of Jyväskylä, Department of Physics,\\
  P.O. Box 35, FI-40014 University of Jyväskylä, Finland}

\affiliation{Helsinki Institute of Physics,\\
  P.O. Box 64, FI-00014 University of Helsinki, Finland}

\emailAdd{petja.k.m.paakkinen@jyu.fi}

\abstract{We present new NLO pQCD predictions for the inclusive photoproduction of dijets in ultra-peripheral (UPC) lead-lead collisions at 5.02 TeV with a realistic impact-parameter dependent effective photon flux obtained through the Woods-Saxon nuclear profile. For the first time in NLO inclusive UPC dijet predictions, we take into account also the modelling of the forward-neutron event class required in the experimental measurements. We show that since the dijet photoproduction at forward rapidities requires an energetic photon in the initial state, this biases the cross section to be dominated by events with relatively small impact parameters between the nuclei, of the order of a few nuclear radii. This leads to a sensitivity to the transverse-plane collision geometry, which we take properly into account by including effects from the finite extent of both the photon-emitting and the target nucleus. We also test the potential sensitivity to the spatial dependence of nuclear parton distribution functions in connection with this finding.}

\FullConference{31st International Workshop on Deep Inelastic Scattering (DIS2024)\\
 8–12 April 2024\\
Grenoble, France\\}

\begin{document}
\maketitle

\vspace{-0.3cm}
\section{Introduction}
\vspace{-0.2cm}

Inclusive photoproduction of dijets in ultra-peripheral nucleus-nucleus collisions (UPCs) has been suggested as a new probe for studying the nuclear parton distributions (nPDFs)~\cite{Strikman:2005yv,Guzey:2018dlm,Guzey:2019kik}. Based on Ref.~\cite{Eskola:2024fhf}, where we used the impact-parameter-dependent equivalent-photon approximation (EPA)~\cite{Baron:1993nk,Greiner:1994db,Krauss:1997vr} for the UPC dijet photoproduction, we show that this process is sensitive to finite-size effects of the incident nuclei and the transverse-plane geometry of the UPC events. It is especially important for the UPC dijets with forward-neutron event class selection~\cite{ATLAS:2022cbd}, which we take into account by including the electromagnetic (e.m.) breakup probability of the photon-emitting nucleus.

\vspace{-0.3cm}
\section{UPC dijets in impact-parameter-dependent EPA with effective photon flux}
\vspace{-0.2cm}

We factorize the inclusive UPC dijet cross section within the impact-parameter-dependent EPA, cf.~Refs.~\cite{Baron:1993nk,Greiner:1994db,Krauss:1997vr},  as
\begin{multline}
    {\rm d} \sigma^{AB \rightarrow A + {\rm dijet} + X} = \sum_{i,j,X'} \int {\rm d}^2{\bf b} \, \Gamma_{AB}({\bf b}) \int {\rm d}^2{\bf r} \, f_{\gamma/A}(y,{\bf r}) \otimes f_{i/\gamma}(x_\gamma,Q^2) \otimes \int {\rm d}^2{\bf s} \, f_{j/B}(x,Q^2,{\bf s}) \\ \otimes {\rm d} \hat{\sigma}^{ij \rightarrow {\rm dijet} + X'}(x_\gamma y p_A,x p_B,Q^2) \times \delta^{(2)}({\bf r} - {\bf s} - {\bf b}),
  \label{eq:xsec_full}
\end{multline}
where $\Gamma_{AB}({\bf b})$ is the probability for no hadronic interaction between the colliding nuclei $A$ and $B$ at an impact-parameter $|{\bf b}|$, $f_{\gamma/A}(y,{\bf r})$ is the equivalent photon flux from nucleus $A$ evaluated at a transverse distance $|{\bf r}|$ from its center, and $f_{i/\gamma}(x_\gamma,Q^2)$ is the photon PDF for a parton $i$ carrying a fraction $x_\gamma$ of the photon energy. Likewise, $f_{j/B}(x,Q^2,{\bf s})$ is the spatially dependent nPDF for a parton $j$ with momentum fraction $x$ at a transverse distance $|{\bf s}|$ from the center of nucleus $B$, and, finally, ${\rm d} \hat{\sigma}^{ij \rightarrow {\rm dijet} + X'}(x_\gamma y p_A,x p_B,Q^2)$ is the cross section for partons $i$ and $j$ (including also the direct photoproduction case where $i=\gamma$) carrying fractions $x_\gamma y$ and $x$ of the per-nucleon beam momenta $p_A$ and $p_B$, respectively, to produce the observed dijet system with a hard scale $Q^2$.

By taking $f_{j/B}(x,Q^2,{\bf s}) = \frac{1}{B} \, T_{B}({\bf s}) \times f_{j/B}(x,Q^2)$, where the nuclear thickness function $T_{B}({\bf s})$ encodes the transverse distribution of partons and $f_{j/B}(x,Q^2)$ is the spatially averaged nPDF, one can simplify the above expression to a form
\begin{equation}
  {\rm d} \sigma^{AB \rightarrow A + {\rm dijet} + X} \!=\!\! \sum_{i,j,X'} f_{\gamma/A}^{\rm eff}(y) \otimes f_{i/\gamma}(x_\gamma,Q^2) \otimes f_{j/B}(x,Q^2) \otimes {\rm d} \hat{\sigma}^{ij \rightarrow {\rm dijet} + X'}(x_\gamma y p_A,x p_B,Q^2),
  \label{eq:xsec_w_eff_flux}
\end{equation}
where we have combined the transverse-space dependence into an \emph{effective} photon flux
\begin{equation}
  f_{\gamma/A}^{\rm eff}(y) = \frac{1}{B} \int {\rm d}^2{\bf r} \int {\rm d}^2{\bf s} \, f_{\gamma/A}(y,{\bf r}) \, T_{B}({\bf s}) \, \Gamma_{AB}({\bf r}\!-\!{\bf s}).
  \label{eq:eff_flux}
\end{equation}
We consider here the following approximations:
\begin{description}
  \item[PL] is the pointlike approximation where the bare photon flux $f_{\gamma/A}^{\rm PL}(y,{\bf r})$ is taken from a point source~\cite{Bertulani:1987tz} and the target nucleus is also treated as a point object, $T_B^{\rm PL}({\bf s}) = B\delta^{(2)}({\bf s})$. Hadronic interactions are excluded with a simple cut-off $\Gamma_{AB}^{\rm PL}({\bf b}) = \theta(|{\bf b}| - 2R_{\rm PL})$, where $R_{\rm PL} = 7.1\ {\rm fm}$ is a hard-sphere radius of a lead nucleus.
  \item[WS$_{\delta({\bf s})}$] refers to an intermediate result where the bare photon flux $f_{\gamma/A}^{\rm WS}(y,{\bf r})$~\cite{Krauss:1997vr}, and the survival probability $\Gamma_{AB}^{\rm WS}({\bf b}) = \exp[-\sigma_{\rm NN} T_{AB}^{\rm WS}({\bf b})]$ with the nuclear overlap function $T_{AB}^{\rm WS}({\bf b})$, are obtained from the Woods-Saxon nuclear distribution, $\sigma_{\rm NN}$ being the total nucleon-nucleon cross section. A delta function for the target distribution $T_B^{\rm PL}({\bf s}) = B\delta^{(2)}({\bf s})$ is still assumed.
  \item[WS] is the full Woods-Saxon approximation, where now also the target nucleus parton spatial distribution is taken from the Woods-Saxon distribution, $T_B^{\rm WS}({\bf s}) = \int_{-\infty}^\infty {\rm d}z \; \rho_B^{\rm WS}\left(\sqrt{z^2+{\bf s}^2}\right)$.
\end{description}

\begin{figure}
  \centering
  \includegraphics[width=0.49725\textwidth]{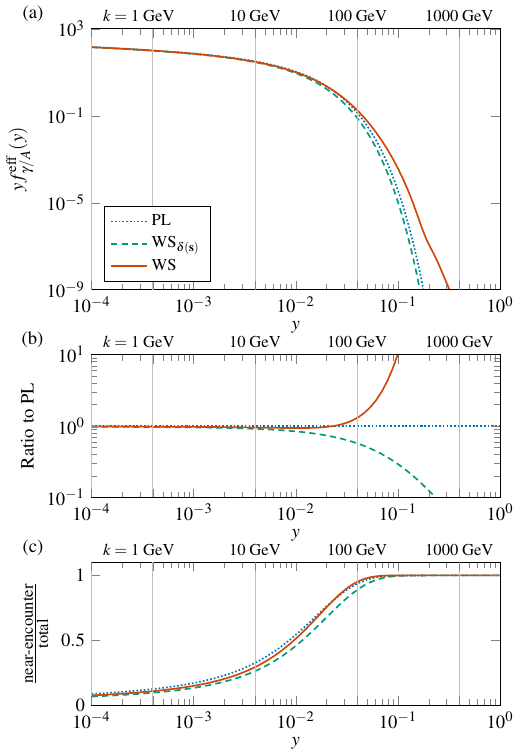}
  \includegraphics[width=0.49725\textwidth]{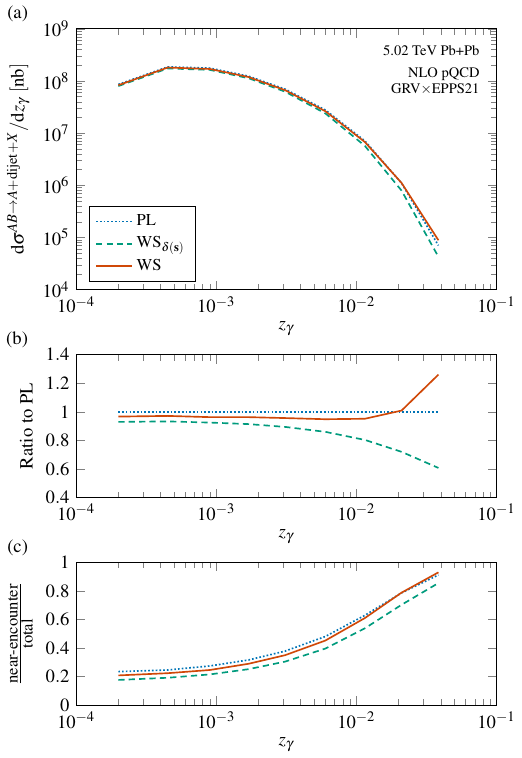}
  \caption{The effective photon flux (left) and UPC dijet cross section (right) in PbPb collisions at $\sqrt{s_{\rm NN}} = 5.02\ {\rm TeV}$, showing the absolute quantities (a), ratio to the pointlike approximation (b), and the fraction of events from $|{\bf r}| < 3R_{\rm PL}$ `near-encounter' configurations (c). Figures from Ref.~\cite{Eskola:2024fhf}.}
  \label{fig:flux-vs-xsec}
  \vspace{-0.25cm}
\end{figure}

The effective fluxes from these approximations and the resulting NLO pQCD UPC dijet cross sections as a function of $z_\gamma = M_{\rm jets} \exp(y_{\rm jets}) / \sqrt{s_{\rm NN}}$, where $M_{\rm jets}$, $y_{\rm jets}$ are the invariant mass and rapidity of the jet system accepted with the ATLAS measurement cuts~\cite{ATLAS:2022cbd}, are compared in Figure~\ref{fig:flux-vs-xsec} for lead-lead collisions at $\sqrt{s_{\rm NN}} = 5.02\ {\rm TeV}$ using GRV~\cite{Gluck:1991jc} and EPPS21~\cite{Eskola:2021nhw} PDFs. We see that for low-energy photons with $k = y\sqrt{s_{\rm NN}}/2 \sim 1\ {\rm GeV}$ (and for small-$z_\gamma$ dijets, correspondingly) the three approximations agree with a good precision, but as the photon energy grows, observable differences emerge. These are caused by the high-energy photons being emitted predominantly in small-impact-parameter `near-encounter' configurations where the transverse-plane event geometry gets resolved~\cite{Eskola:2024fhf}. In particular, we note that the full WS prediction at large $z_\gamma$ differs from both the PL and WS$_{\delta({\bf s})}$ approximations, showing that it is necessary to take into account the finite-size effects both in the survival probability $\Gamma_{AB}({\bf b})$ and the target-nucleus spatial distribution $T_{B}({\bf s})$.

\vspace{-0.3cm}
\section{Modelling of the neutron-class event selection and spatially dependent nPDFs}
\vspace{-0.2cm}

\begin{figure}
  \centering
  \raisebox{-\height}{\includegraphics[width=0.49725\textwidth]{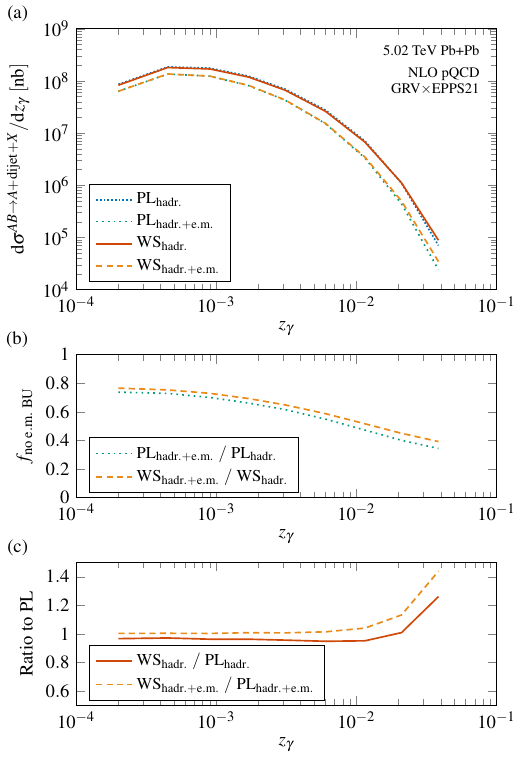}}
  \raisebox{-\height}{\includegraphics[width=0.49725\textwidth]{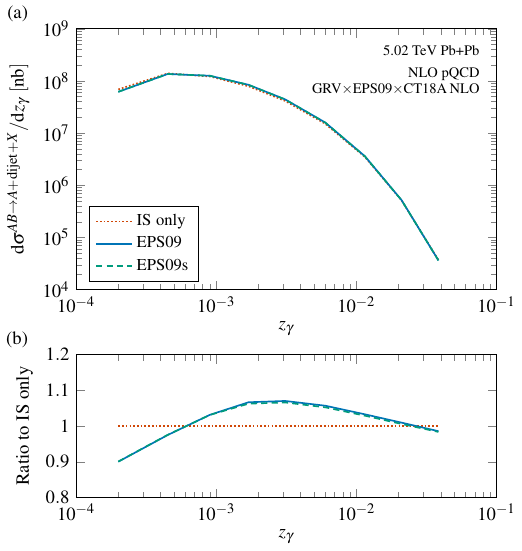}}
  \caption{The impact of e.m.\ breakup modelling (left) and spatial dependence of nuclear modifications on the UPC dijet cross section, showing absolute quantities (a) and various ratios (b, c). Figures from Ref.~\cite{Eskola:2024fhf}.}
  \label{fig:xsec-em-bu-and-EPS09s}
  \vspace{-0.25cm}
\end{figure}

The results shown in Figure~\ref{fig:flux-vs-xsec} were fully inclusive, thus allowing for any number of forward neutrons in either beam direction. To determine which of the two colliding ions in UPCs serves as a nuclear target, one can use zero-degree calorimeters and perform the measurements in the $0nXn$ event class~\cite{ATLAS:2022cbd}. For the requirement of zero neutrons in the photon-going direction, we include the e.m.\ breakup probability from Starlight~\cite{Klein:2016yzr} in the survival factor $\Gamma_{AB}({\bf b})$. The impact is shown in Figure~\ref{fig:xsec-em-bu-and-EPS09s} (left), where we see that this leads to a large reduction in the cross section, particularly at large $z_\gamma$, where it also increases the difference between the WS and PL approximations. For having $X \geq 1$ neutrons in the opposite direction, we should still subtract a diffractive contribution from our result, but this is expected to be a small correction~\cite{Eskola:2024fhf}.

We have also studied exchanging the simple factorization assumption $f_{j/B}(x,Q^2,{\bf s}) = \frac{1}{B} \, T_{B}({\bf s}) \times f_{j/B}(x,Q^2)$ with a more realistic phenomenological parametrization from EPS09s~\cite{Helenius:2012wd}. The result is shown in Figure~\ref{fig:xsec-em-bu-and-EPS09s} (right), comparing the spatially dependent EPS09s nuclear modifications to spatially independent EPS09 nuclear modifications and to predictions with the trivial isospin (IS) effects only. We find that the spatially dependent nuclear modifications give only a small correction to the spatially averaged ones in the cross section, and thus the simple factorization with WS transverse profile captures the main spatial effects.

\vspace{-0.3cm}
\section{Conclusions}
\vspace{-0.2cm}

We have shown that the UPC dijet production at large $z_\gamma$ is sensitive to the assumptions on the nuclear transverse profile and collision geometry in `near-encounter' events, and thus a realistic calculation of the effective photon flux is needed for an accurate interpretation of the measurement. Furthermore, taking into account the nuclear breakup through additional e.m.\ interactions is necessary when comparing with data in the $0nXn$ event class, but the spatial resolution of this process appears not to be detailed enough for constraining the impact-parameter dependence of the nPDF nuclear modifications.

\vspace{-0.2cm}
\acknowledgments
\vspace{-0.2cm}
This research was funded through the Research Council of Finland projects No.\;330448 and No.\;331545, as a part of the Center of Excellence in Quark Matter of the Research Council of Finland (projects No.\;346325, No.\;346326, No.\;364192 and No.\;364194) and as a part of the European Research Council project ERC-2018-ADG-835105 YoctoLHC. We acknowledge computing resources from the Finnish IT Center for Science (CSC), utilised under the project jyy2580.

\end{document}